# Combining Metadynamics and Integrated Tempering Sampling

Yi Isaac Yang[1,2], Haiyang Niu[1,2] and Michele Parrinello[1,2,3,*]

1. *Department of Chemistry and Applied Biosciences, ETH Zurich, c/o USI Campus, Via Giuseppe Buffi 13, CH-6900, Lugano, Ticino, Switzerland*
2. *Institute of Computational Science, Università della Svizzera italiana (USI), Via Giuseppe Buffi 13, CH-6900, Lugano, Ticino, Switzerland*
3. *Instituto Italiano di Tecnologia,Via Morego 30, 16163 Genova*
* Correspondence to: parrinello@phys.chem.ethz.ch

## Abstract

The simulation of rare events is one of the key problems in atomistic simulations. Towards its solution a plethora of methods have been proposed. Here we combine two such methods metadynamics and integrated tempering sampling. In metadynamics the fluctuations of a carefully chosen collective variable are amplified, while in integrated tempering sampling the system is pushed to visit an approximately uniform interval of energies and allows exploring a range of temperatures in a single run. We describe our approach and apply it to the two prototypical systems a $S_N2$ chemical reaction and to the freezing of silica. The combination of metadynamics and integrated tempering sampling leads to a powerful method. In particular in the case of silica we have measured more than one order of magnitude acceleration.

## Introduction

While atomistic simulations have become very widely used in many fields of science, still they suffer from many limitations. Such limitations hamper an even wider range of applications. Particularly problematic is the limited time scale that can be simulated. In order to address this issue many enhanced methods have been suggested. For a recent review, we refer the reader to a recent book[1].

Without having the pretence of being rigorous or exhaustive, enhanced sampling methods can be classified into two categories, according to whether they use collective variables (CVs) or not. Among those that do not we can count parallel tempering[2], Wang-Landau[3], accelerated molecular dynamics[4], and integrated tempering sampling (ITS)[5,6]. Among the former some of the names that spring to mind are, umbrella sampling[7], metadynamics (MetaD)[8,9], and variationally enhanced sampling (VES)[10] . The two kinds of approaches have well known strengths and weaknesses that are often complementary. It is therefore profitable to try and combine methods that come from the two classes. Some examples of this hybrid approach are the combination of parallel tempering and metadynamics[11] or the combination of ITS and umbrella sampling[12].

Here we want to combine ITS with metadynamics. Such a method recommends itself for a number of features. Contrary to parallel tempering its computational cost is negligible, and reweighting relatively easy allowing the properties at different temperature to be calculated at no additional cost. The method acts by enhancing energy fluctuations so as to cover a pre-assigned range of energies and gives an amount of control larger than what is offered by, say, the well-tempered ensemble[13,14]. However, being generic it can fail to target the degrees of freedom that are relevant to the specific rare event of interest. This can be more easily done in metadynamics as already realized in ref[15], where the two schemes where made to work on the same problem at different times. Here we bring the two methods to work together in an integrated manner as a part of the same algorithm. The end result is a very efficient algorithm that very often has a superior rate of convergence than either metadynamics or ITS on their own. In addition the use of ITS allows calculating the property of the system at different temperatures at the cost of one single simulation.

We first present the theoretical underpinnings of what we do and then we apply the method to an $S_N2$ chemical reaction and to the solid-liquid phase transition of silica.

## Methodology

We recall some of the main features of ITS. However, we warn the reader familiar with ref[5,16] that here the method is looked at from a different point of view and that our notation is different.

To begin with, we observe that the potential $U(\mathbf{R})$ that is used to drive an MD simulation can be rewritten, modulo an irrelevant constant, as

$$U_{\text{eff}}(\mathbf{R}) = -\frac{1}{\beta} \log P_\beta(\mathbf{R}) \quad (1)$$

where $P_\beta(\mathbf{R}) = 1/Z_\beta \exp[-\beta U(\mathbf{R})]$ with $Z_\beta = \int d\mathbf{R} e^{-\beta U(\mathbf{R})}$ is the normalized Boltzmann distribution. Parenthetically we recall here that the free energy $F_\beta$ can be obtained from the partition function $Z_\beta$ as $F_\beta = -1/\beta \log Z_\beta$. In an attempt to enhance energy fluctuations in a controlled way and in a spirit similar to that of parallel tempering[2], one substitutes $P(\mathbf{R})$ in equation (1) with an average over $N$ normalized Boltzmann distributions at different temperatures. Thus $P_\beta(\mathbf{R})$ is replaced by $1/N \sum_{k=1}^{N} P_{\beta_k}(\mathbf{R})$ and $U_{\text{eff}}(\mathbf{R})$ becomes $U_{\text{eff}}(\mathbf{R}) = -1/\beta \log[1/N \sum_{k=1}^{N} P_{\beta_k}(\mathbf{R})]$, or equivalently:

$$U_{\text{eff}}(\mathbf{R}) = -\frac{1}{\beta} \log \left[ \frac{1}{N} \sum_{k=1}^{N} \frac{1}{Z_{\beta_k}} e^{-\beta_k U(\mathbf{R})} \right] \quad (2)$$

Use of this expression however implies that we know the normalizing partition functions $Z_{\beta_k}$, that is to say that we know the free energy of the system at each temperature $\beta_k$. This is a daunting task, however, Gao in his ITS method[16] has been able to recast the problem into that of calculating partition functions ratios between neighbouring $\beta_k$'s. Of course this is a much simpler undertaking and can be made even simpler by taking small $\beta_{k-1} - \beta_k$ intervals. This reformulation is at the heart of the success of ITS. However, beforehand a few manipulations are needed.

First, we notice that to all intense and purpose $U_{\text{eff}}(\mathbf{R})$ equation (2) can replaced by

$$U'_{\text{eff}}(\mathbf{R}) = -\frac{1}{\beta}\log\left[\frac{1}{N}\sum_{k=1}^{N}\frac{Z_{\beta_1}}{Z_{\beta_k}}e^{-\beta_k U(\mathbf{R})}\right] \quad (3)$$

since $U_{\text{eff}}(\mathbf{R})$ differs from $U'_{\text{eff}}(\mathbf{R})$ only by the constant $1/\beta \log(NZ_{\beta_1})$. Each term in the sum that defines $U'_{\text{eff}}(\mathbf{R})$ is no longer normalized to one but still they all will be normalized to the same value.

The next step is to note that the partition functions ratios $Z_{\beta_1}/Z_{\beta_k}$ in $U'_{\text{eff}}(\mathbf{R})$ can be written as $Z_{\beta_1}/Z_{\beta_k} = \prod_{l=1}^{l=k}[Z_{\beta_{l-1}}/Z_{\beta_l}]$. This implies that, once the ratios $Z_{\beta_{k-1}}/Z_{\beta_k}$ are known, $U'_{\text{eff}}(\mathbf{R})$ can be reconstructed. Each ratio can be written as $Z_{\beta_{k-1}}/Z_{\beta_k} = \int d\mathbf{R}e^{-\beta_{k-1}U(\mathbf{R})} / \int d\mathbf{R}e^{-\beta_k U(\mathbf{R})}$, a quantity that in turn can be computed in a run driven by $U'_{\text{eff}}(\mathbf{R})$ at the temperature $\beta$ using the identity

$$\frac{Z_{\beta_{k-1}}}{Z_{\beta_k}} = \frac{\int d\mathbf{R}e^{-\beta_{k-1}U(\mathbf{R})}}{\int d\mathbf{R}e^{-\beta_k U(\mathbf{R})}} = \frac{\left\langle e^{-\beta_{k-1}\left[U(\mathbf{R})-\frac{\beta}{\beta_{k-1}}U'_{\text{eff}}(\mathbf{R})\right]}\right\rangle_{U'_{\text{eff}}(\mathbf{R})}}{\left\langle e^{-\beta_k\left[U(\mathbf{R})-\frac{\beta}{\beta_k}U'_{\text{eff}}(\mathbf{R})\right]}\right\rangle_{U'_{\text{eff}}(\mathbf{R})}} = \frac{\left\langle e^{\beta_{k-1}V_{k-1}^{\text{ITS}}[U(\mathbf{R})]}\right\rangle_{U'_{\text{eff}}(\mathbf{R})}}{\left\langle e^{\beta_k V_k^{\text{ITS}}[U(\mathbf{R})]}\right\rangle_{U'_{\text{eff}}(\mathbf{R})}} \quad (4)$$

where the averages $\langle\cdots\rangle_{U'_{\text{eff}}(\mathbf{R})}$ are calculated in the ensemble generated by $U'_{\text{eff}}(\mathbf{R})$ and we have defined the effective bias potential $V_k[U(\mathbf{R})]$ acting at temperature $\beta_k$ as:

$$V_k[U(\mathbf{R})] = \frac{\beta}{\beta_k}U'_{\text{eff}}(\mathbf{R}) - U(\mathbf{R}) = -\frac{1}{\beta_k}\log\left[\frac{1}{N}\sum_{k=1}^{N}\frac{Z_{\beta_1}}{Z_{\beta_k}}e^{-\beta_k U(\mathbf{R})}\right] - U(\mathbf{R}) \quad (5)$$

This set of self-consistent equations define ITS and can be solved with an iterative procedure. One starts with a guess for $\frac{Z_{\beta_{k-1}}}{Z_{\beta_k}}$, after some convenient interval of time $\tau_I$ the estimation of $Z_{\beta_{k-1}}/Z_{\beta_k}$ is updated using equation (4) the driving potential $U'_{\text{eff}}(\mathbf{R})$ is changed accordingly and the procedure is iterated. A prescription to accelerate reaching the stationary condition is given in ref[16]. Once the stationary situation is reached, one stops the self-consistency procedure and use the last $U'_{\text{eff}}(\mathbf{R})$ in a standard umbrella-like reweighting mode, see equation (8) below.

One advantage of using the ITS approach is that the computational overhead of running the system under the action of $U'_{\text{eff}}(R)$ is negligible, in fact the equation of motions read:

$$M\ddot{R} = f[U(R)] \cdot \left[-\frac{\partial U(R)}{\partial R}\right] = f[U(R)]F(R) \tag{6}$$

with

$$f[U(R)] = \frac{\sum_{k=1}^{N}\left[\beta_k \frac{Z_{\beta_1}}{Z_{\beta_k}} e^{-\beta_k U(R)}\right]}{\beta \sum_{k=1}^{N}\left[\frac{Z_{\beta_1}}{Z_{\beta_k}} e^{-\beta_k U(R)}\right]} \tag{7}$$

Thus, one only needs to calculate $f[U(R)]$ and rescale the force $F(R)$ accordingly, at each time step. One could read equation (6) as if the effect of using $U'_{\text{eff}}(R)$ is simply to modulate the mass $M$ of the particles according to the value of the energy. This is reminiscent of the accelerated molecular dynamics method[4] where also one changes the mass depending on the value of $U$ but here is done differently and in a different spirit.

Any enhanced sampling method is not complete unless it is accompanied by an appropriate reweighing method. Namely, there has to be a way of calculating the Boltzmann's equilibrium average of any operator from the biased run. This is all the more relevant in the case of ITS since it promises to yield the properties of the systems at all the $N$ temperatures $\beta_k$ in one single run. A straightforward application of umbrella-like reweighting gives for the average of an operator $O(R)$ at temperature $\beta_k$ the expression:

$$\langle O(R)\rangle_{\beta_k} = \frac{\langle O(R)e^{\beta_k V_k[U(R)]}\rangle_{U'_{\text{eff}}(R)}}{\langle e^{\beta_k V_k[U(R)]}\rangle_{U'_{\text{eff}}(R)}} \tag{8}$$

The combination of ITS and metadynamics is now ready to be introduced in a natural way and it simply amounts at introducing in $U'_{\text{eff}}(R)$ a history dependent bias $V'[s(R), t]$ that is a function of an appropriately selected set of collective variables $s(R)$:

$$U_{\text{eff}}(\boldsymbol{R},t) = -\frac{1}{\beta}\log\left[\frac{1}{N}\sum_{k=1}^{N}\frac{1}{Z_{\beta_k}}e^{-\beta_k\{U(\boldsymbol{R})+V'[s(\boldsymbol{R}),t]\}}\right] \tag{9}$$

The updating rules for $V'[s(\boldsymbol{R}),t]$ are the same as for standard metadynamics, namely at regular interval of times $\tau_M$ the bias is modified by adding a repulsive Gaussian at the current location $s[\boldsymbol{R}(t)]$ a repulsive Gaussian $w_g \exp\left[-\beta \frac{V(s',t)}{\gamma-1}\right]\exp\left\{-\beta\frac{[s(\boldsymbol{R})-s']^2}{2\sigma_g^2}\right\}$, where $\sigma_g$ is the Gaussians width, $w_g$ their initial height and $\gamma$ the well-tempered metadynamics boosting parameter. The value of these parameters will be chosen following standard metadynamics prescriptions. It seems natural at this stage to update synchronously both the bias and the partition function ratio at the same time interval $\tau = \tau_I = \tau_M$. Other choices are possible but will not be explored here. Also in this case we first reach a stationary condition and then freeze the potential and evolve the system with this fixed bias using an umbrella sampling reweighting to calculate the equilibrium averages.

## Results

In order to illustrate the performance of this new hybrid method, that we shall call MetaITS, we have chosen two applications taken from different areas. The first is a typical $S_N2$ reaction and the second is the crystallization of silica. In these two examples the performance of ITS is poor, albeit for different reasons. However, if ITS is supplemented with metadynamics using the MetaITS protocol the calculations can be satisfactorily converged. Particularly remarkable is the case of silica where one gains almost one order of magnitude speed up in convergence not to mention the fact that in all cases one gets information on all the $N$ $\beta_k$ temperature in one single simulation.

## S_N2 Chemical Reaction

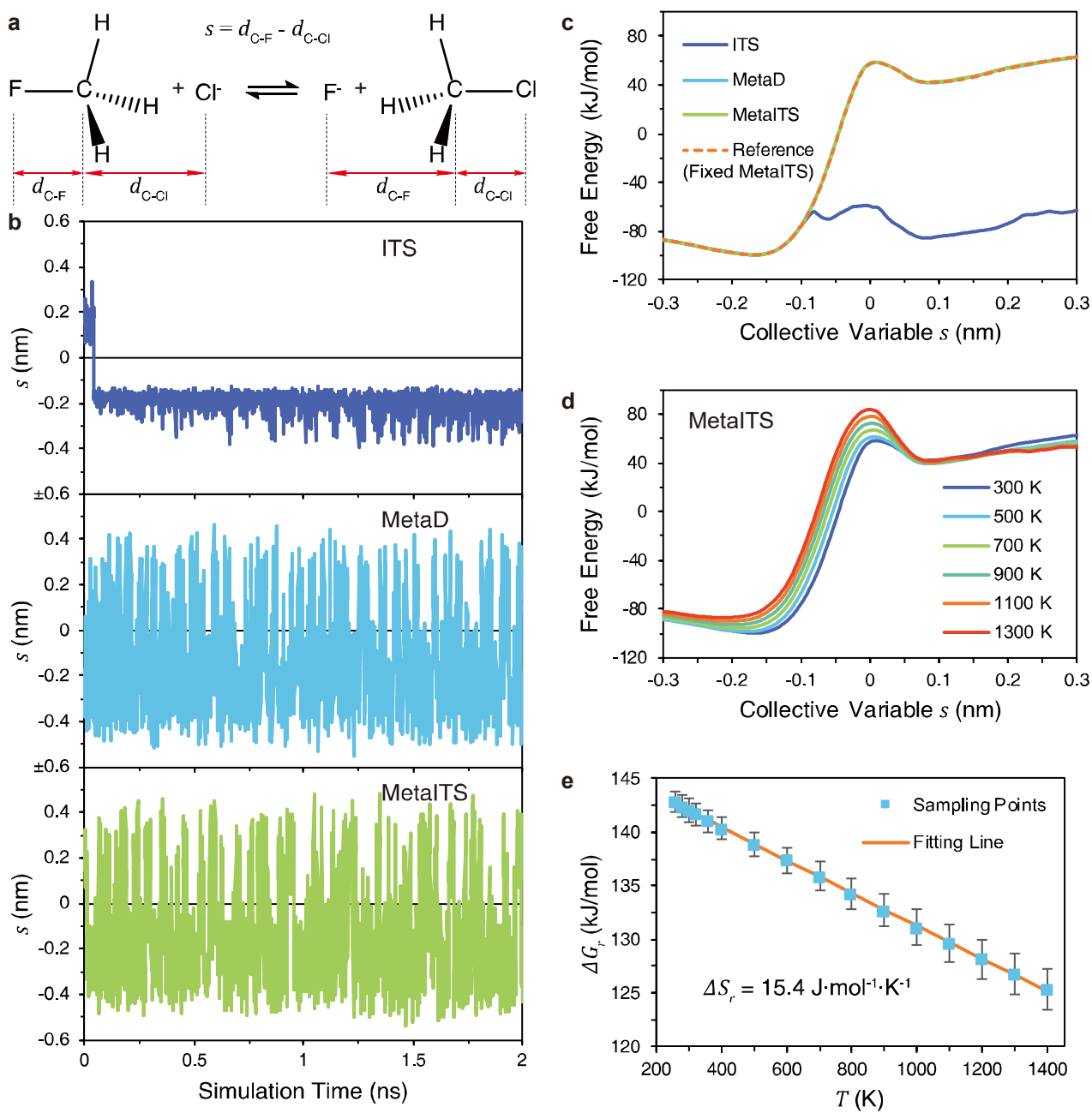

**Figure 1** a) The scheme of the S$_N$2 reaction and the definition of the CV b) Evolution of the CV $s$ as the function of time using the three enhanced sampling methods ITS, MetaD and MetaITS. c) Free energy surface as the function of CV $s$ at 300 K calculated from the simulation using different methods. d) Variation of the free energy surface profile as function of $s$ at different temperatures as computed from MetaITS. e) Free energy difference $\Delta G_r$ from reactant (CH$_3$F) to product (CH$_3$Cl) at different temperatures and the fitting line of the $\Delta G_r$ as the function of temperature. Using the slope of the fitted line, we can estimate the value of reaction entropy $\Delta S_r$ is about 15.4 J·mol$^{-1}$·K$^{-1}$.

We first tested the method on a simple S$_N$2 reaction at vacuum using a quantum mechanical empirical model to describe the interatomic forces. The reaction studied is $CH_3F + Cl^- \rightleftharpoons CH_3Cl + F^-$, whose reaction scheme is shown at Figure 1a. We defined the distance between C atom and F atom as $d_{C-F}$ and the distance between C atom and Cl atom as $d_{C-Cl}$. Following the work of Piccini *et al.*[17], in both the MetaD and MetaITS runs, we used the difference between these two distance $s = d_{C-F} - d_{C-Cl}$ as CV. In ITS and MetaITS, we split the inverse temperature range 1/250 - 1/1500 K$^{-1}$ into $N = 360$ intervals. With this choice we performed MD simulations at 300 K with the three enhanced sampling methods ITS, metadynamics and MetaITS. Each simulation was run for 2 ns.

The evolution of the CV as a function of time are compared in Figure 1b. In this case the ITS method is rather ineffective since the chemical reaction only makes one transition from $CH_3Cl$ to $CH_3F$ and is not able to go back within the time of the simulation. In contrast, both MetaD and MetaITS could drive the chemical reaction forth and back very frequently. The results of the free energy surfaces of MetaD and MetaITS are compatible with one another (Figure 1c) and the methods appear to be similarly efficient. We also used the converged coefficients of MetaITS to perform the simulation with a fixed bias potential $V(s)$ for 5 ns and used the resulting FES as reference. It shows that the both of the FES of MetaD and MetaITS are equal to the reference one. However, using MetaITS, one has the additional bonus of getting the properties of the system at all the $\beta_k^{-1}$ temperatures in one single run. This allows calculating other interesting thermodynamic properties, like entropy. For example, Figure 1e shows the free energy difference $\Delta G_r$ of the transition from reactant ($CH_3F$) to product ($CH_3Cl$) at different temperatures. From this one can calculate the reaction entropy $\Delta S_r$ from the relation $\Delta S_r = -\left(\frac{\partial \Delta G_r}{\partial T}\right)_p$. After fitting the sampled points with a straight line, the value of $\Delta S_r$ is estimated to be about 15.4 J·mol$^{-1}$·K$^{-1}$. We find remarkable that in one simulation we can compute quantities like entropy that usually require performing several simulations.

**Liquid Silica Crystallization**

Next, we test the method with a more complex and physically more relevant system, liquid silica crystallization. Crystallization is a typical first-order phase transition in which the disordered atoms of a liquid spontaneously form ordered periodic patterns. Understanding the way in which crystallization proceeds also holds the key to improving many scientific and technological processes. In a recent publication[18] we were able to identify an excellent CV and were able to study this process. Many difficulties, such as the high crystallization barrier and the high tendency of liquid silica to form glasses, had to be overcome and MetaD played a crucial role. However in order to estimate the free energy surface (FES), melting temperature and entropy difference between solid and liquid, several independent metadynamics simulations had to be performed at different temperatures. Furthermore, the transition rate between solid and liquid in the biased runs depends significantly on temperature. For instance the transition rate at 2,300 K is more than 10 times slower than that at 2,800 K, which makes the convergence of the FES at low temperatures time-consuming. As we shall see below MetaITS was able to solve all these problems in a single simulation.

Also here we perform three different simulations with the three different methods, ITS, MetaD, and MetaITS. In the ITS simulations (ITS and MetaITS), the temperature interval 2,100 to 3,000 K was split in $N$=300 intervals. The melting temperature of $\beta$-cristobalite (about 2,350 K[18]) is included in this interval but not at its centre, since taking a high proportion of temperature intervals above melting is expected to accelerate convergence as we shall discuss later. In the MetaD run the intensity of the {111} peak of the X-ray diffraction pattern of $\beta$-cristobalite was taken as CV[18]. To favour comparison all the other parameters are kept equal in all the simulations.

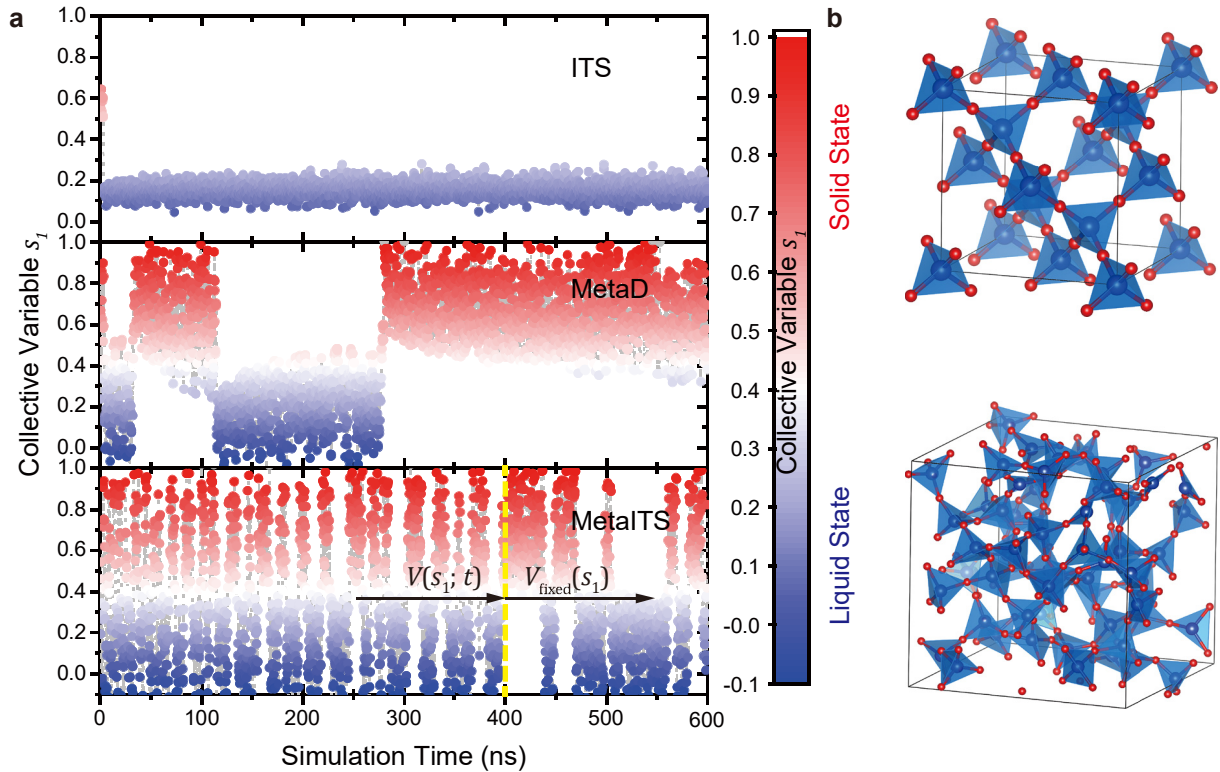

**Figure 2** a) Collective variable $s_1$ as a function of simulation time at 2,350 K by ITS, MetaD, and MetaITS in the application to liquid silica crystallization. The collective variable $s_1$ refers to the X-ray diffraction (XRD) intensity of the {111} peak of $\beta$-cristobalite, which is rescaled by the intensity value (~242) of ideal $\beta$-cristobalite. Due to the nature of Debye scattering equation for calculating the XRD pattern, negative intensity value can be seen. Only one melting event from $\beta$-cristobalite (red) to liquid (blue) is observed in ITS simulation, while reversible transitions between $\beta$-cristobalite (red) and liquid (blue) are obtained in MetaD and MetaITS simulations. After the calculation of MetaITS reached convergence, we fixed the bias potential $V(s_1; t)$ at 400 ns to continue the calculation for 200 ns. b) The structure models of liquid and solid ($\beta$-cristobalite) silica phases.

Figure 2 contrast vividly the performance of the three methods. On its own ITS fails miserably, once the system melted the liquid was not able to recrystallize. In contrast a number of transitions between solid and liquid phases can be observed when using standard MetaD. However it has to be observed that the relative low transition frequency makes convergence slow. A remarkable changes takes place when we combine MetaD and ITS. The phase transition frequency is increased by at least a factor of 10. One of the reason for this remarkable acceleration is that close to melting $SiO_2$ is very viscous. By enhancing fluctuation towards more energetic and fluid states a more raid diffusion in configuration space is favoured. This is the reason why in choosing the $\beta_k$ interval, we have extended the $\beta_k$ intervals more in the high temperature regions.

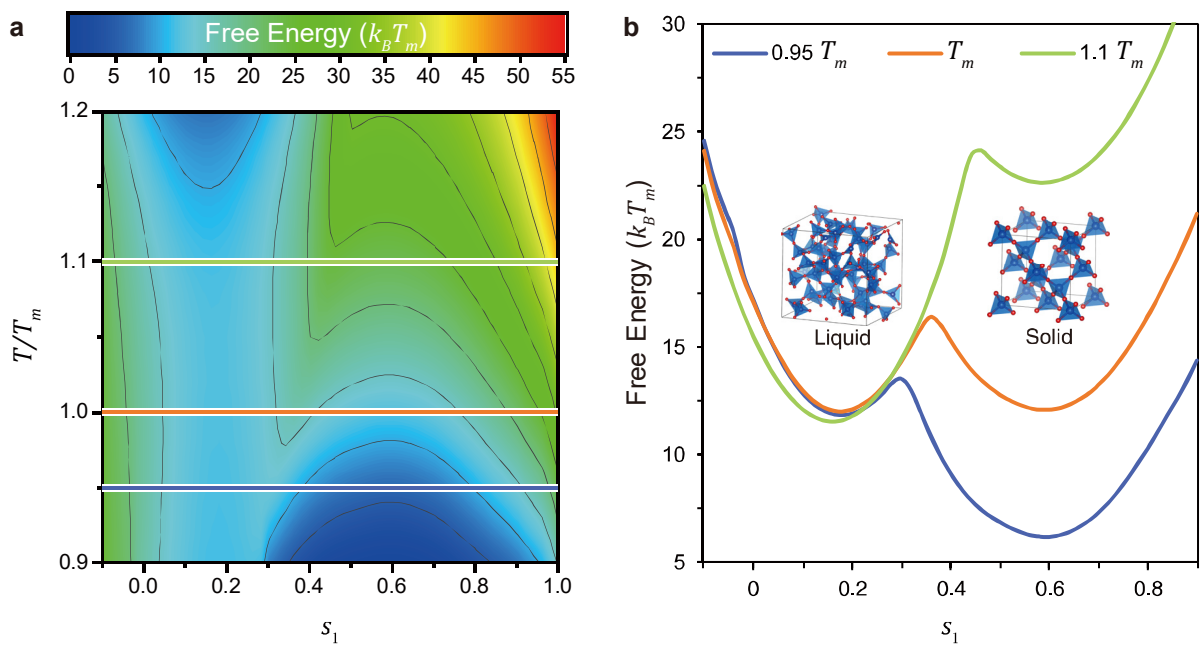

**Figure 3** a) Free energy surface (FES) in terms of collective variable s1 and temperature. b) FES as the function of s1 at three temperatures, $T < T_m$, $T = T_m$, and $T > T_m$. The left and right basins refer to the liquid and solid phase of silica, respectively. The structure models of liquid and solid silica phases are shown in the corresponding basin. The collective variable s1 refers to the intensity of the {111} peak of β-cristobalite, which is rescaled by the intensity value (~242) of ideal β-cristobalite.

Of course this high transition rate accelerates the convergence of the calculation. As shown in Figure 2a, after fixing the bias potential $V_{\text{fixed}}(s_1)$ several reversible transitions between solid and liquid were seen. Therefore, accurate FES can be reweighted. In Figure 3 we show the FES as a function of the MetaD CV $s_1$ and temperature. This is made possible by the magic of ITS that allows in a single run to compute the properties at different temperatures. This is to be contrasted with the MetaD calculation of ref [18] where for each temperature a different MetaD run was necessary. If we compound this advantage with the increased convergence rate the advantages of using MetaITS are really amazing.

Within statistical errors the results are in agreement with those of Ref[18]. Also here we can use the information contained in temperature dependence of the FES to obtain the properties of the liquid to solid transition. One first computes the free energy difference $\Delta G_{S \to L}$ between liquid and solid as a function of temperature. The result are shown in Figure 4 and compared with those of ref [18]. As can be seen, the results with these two methods are in agreement with one another. The melting temperature calculated from the intercept of the $\Delta G_{S \to L}$ axis with the zero gives $T_m$ = 2,334 ± 6 K, in agreement with the MetaD result (2,342 ± 15 K). Similarly the difference in entropy between $\beta$-cristobalite and liquid, $\Delta S_{S \to L}$, can be calculated from the thermodynamic relation $\Delta S_{S \to L} = -\left(\frac{\partial \Delta G_{S \to L}}{\partial T}\right)_{N,P}$. The estimated $\Delta S_{S \to L}$ at melting temperature is about 14.6 J·K$^{-1}$·mol$^{-1}$, once again in agreement with the result (14.6 J·K$^{-1}$·mol$^{-1}$) obtained by MetaD.

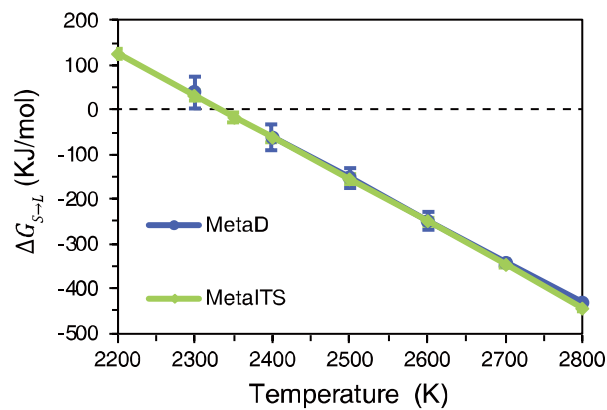

**Figure 4** Difference in free energy $\Delta G_{S \to L}$ between $\beta$-cristobalite and liquid phases as a function of temperature obtained by MetaD and MetaITS.

# Conclusion

The results shown here show how the synergetic effect of merging MetaD and ITS leads to a method in which the strength of the two method are combined in a non-linear fashion and their weaknesses are lifted. The case of Silica illustrate well the situations in which meta ITS is expected to work at is best, that is a transition between a disordered and an ordered state in which the ordered state occupied a small fraction of the available configuration states and this ordered state cannot be discovered only by enhancing the energy fluctuations. The use of metadynamics with an appropriately defined order parameter helps finding the ordered state while the ITS accelerate sampling in the two states. We believe however that much needs to be done and considerable room for improvement do exist. Also the combination of the two methods gives a very useful insight into how the two methods work.

## Methods

### Simulation detail

For the system of the $S_N2$ reaction, the MD simulation were performed using SANDER program at AmberTools 17 software package[19]. The simulations are been performed using the QM/MM model of SANDER, and all the atoms of the system are chosen as the QM part with semi-empirical model PM6[20]. No constrain is put to chemical bond distances, and no cut-off is used for the non-bonding interactions. A time step of 0.5 fs is utilized. For all the simulation, harmonic restraining walls have been added on $d_{C-F}$ and $d_{C-F}$ for distances larger than 0.5 nm to prevent the atoms moving too far away, which is utilized using the bias UPPER_WALL in the PLUMED2 plug-in library[21] (KAPPA is set to 150 kJ/mol). For ITS simulation[5], the temperature range is from 250 K to 1500 K with 300 intervals. The coefficients $Z_{\beta_1}/Z_{\beta_k}$ were updated every 200 steps (0.1 ps). For metadynamics simulation[8], the height and width of Gaussians are 10 kJ/mol and 0.012 nm, respectively, and are deposited every 100 steps (0.05 ps). The bias factor $\gamma$ of well-tempered metadynamics[22] is set to 50.

For silica, we performed an isothermal–isobaric molecular dynamics (MD) simulation. All MD simulations were performed with large-scale atomic/molecular massively parallel simulator (LAMMPS)[23]. The integration of the equations of motion was carried out with a time step of 2 fs. We used the stochastic velocity rescaling thermostat with a relaxation time of 0.1 ps[24]. The target pressure of the Parrinello–Rahman barostat[25] was set to the standard atmosphere value and a relaxation time of 10 ps was used. We used the interatomic potentials reported in ref.[26] for our MD simulations. To determine the free-energy surface (FES) $G(s_1)$ as a function of CV, we used well-tempered metadynamics (WTMetaD)[22]. We set the bias factor of the WTMetaD ensemble equal to the value 100. The WTMetaD bias in potential space was constructed by depositing Gaussians every 1 ps with width 5 CV units and a height of 40 kJ/mol.

**Code availability**

In this work, the enhanced sampling functions (metadynamics and ITS) are realized by a modified PLUMED2 plug-in library based on the official version 2.4 (http://www.plumed.org/). The modified version is added the function of the ITS method, which is available free of charge: https://github.com/helloy-esterday/plumed_its_full.

# Acknowledgements

The authors thank Yi Qin Gao, Lijiang Yang, Jun Zhang and GiovanniMaria Piccini for useful discussion. Computational resources were provided by the Swiss National Supercomputing Centre (CSCS). This research was supported by the VARMET European Union Grant ERC-2014-ADG-670227 and the NCCR MARVEL 51NF40_141828 funded by the Swiss National Science Foundation.

## Author contributions

Y.I.Y., H.N. and M.P. designed research and wrote the paper. Y.I.Y. and M.P. developed methods. Y.I.Y. and H.N. performed/analysed simulations. Y.I.Y. developed software.

## Competing financial interests

The authors declare no competing financial interests.

# A Simple Guide to use MetaITS with PLUMED2

## Supplementary material: Combining Metadynamics and Integrated Tempering Sampling

Yi Isaac Yang[1,2], Haiyang Niu[1,2] and Michele Parrinello[1,2]

1. *Department of Chemistry and Applied Biosciences, ETH Zurich, c/o USI Campus, Via Giuseppe Buffi 13, CH-6900, Lugano, Ticino, Switzerland*
2. *Institute of Computational Science, Università della Svizzera italiana (USI), Via Giuseppe Buffi 13, CH-6900, Lugano, Ticino, Switzerland*
3. *Instituto Italiano di Tecnologia,Via Morego 30, 16163 Genova*
* Correspondence to: parrinello@phys.chem.ethz.ch

**Table of Content: A Simple Guide to use MetaITS with PLUMED2**

# S-I  To perform ITS using PLUMED2

Here is an example of the PLUMED2 input file to perform ITS:

```
# Potential energy of the simulation system
energy: ENERGY

# The module of ITS
ITS_BIAS ...
  LABEL=its
# Set potential energy as the CV of ITS
  ARG=energy
# The number of replicas
  NREPLICA=100
  SIM_TEMP=300
# The temperature range
  TEMP_MAX=370
  TEMP_MIN=270
# The update step
  PACE=200
# The temperatures need to reweight
  RW_TEMP=280,300,320,340,360
... ITS_BIAS

PRINT ...
  ARG=energy,its.*
  FILE=colvar.bias.data
  STRIDE=100
... PRINT
```

Some important terms at the PLUMED2 output file "colvar.bias.data" are:

- its.bias: the original bias potential ($V\{U[\boldsymbol{R}(t)]\}$) of ITS at the simulation temperature
- its.rct: the revise factor $c(t)$ of ITS at the simulation temperature
- its.rbias: the revised bias potential ($V'\{U[\boldsymbol{R}(t)]\} = V\{U[\boldsymbol{R}(t)]\} - c(t)$) of ITS at the simulation temperature
- its.rbias_T280, its.rbias_T300, its.rbias_T320, its.rbias_T340, its.rbias_T360: the revised bias potential at different temperatures that you want to reweight (set at RW_TEMP)

## S-II    Reweighting of ITS

Here is an example of the PLUMED2 input file to calculate the histogram of this value at temperatures 280 K, 300 K and 320 K:

```
cv: READ FILE=colvar.info.data VALUES=CV IGNORE_TIME
its: READ FILE=colvar.bias.data VALUES=its.* IGNORE_TIME

rw280: REWEIGHT_BIAS ARG=its.rbias_T280 TEMP=280
rw300: REWEIGHT_BIAS ARG=its.rbias_T300 TEMP=300
rw320: REWEIGHT_BIAS ARG=its.rbias_T320 TEMP=320

HISTOGRAM ...
  LABEL=h280
  ARG=cv
  GRID_MIN=0.0
  GRID_MAX=1.0
  GRID_BIN=100
  BANDWIDTH=0.2
  LOGWEIGHTS=rw280
... HISTOGRAM

HISTOGRAM ...
  LABEL=h300
  ARG=cv
  GRID_MIN=0.0
  GRID_MAX=1.0
  GRID_BIN=100
  BANDWIDTH=0.2
  LOGWEIGHTS=rw300
... HISTOGRAM

HISTOGRAM ...
  LABEL=h320
  ARG=cv
  GRID_MIN=0.0
  GRID_MAX=1.0
  GRID_BIN=100
  BANDWIDTH=0.2
  LOGWEIGHTS=rw320
... HISTOGRAM

DUMPGRID GRID=h280 FILE=histo.t280.data
DUMPGRID GRID=h300 FILE=histo.t300.data
DUMPGRID GRID=h320 FILE=histo.t320.data
```

## S-III    To perform MetaITS using PLUMED2

Here is an example to perform MetaITS:

```
energy: ENERGY
d1: DISTANCE ATOMS=3,5
d2: DISTANCE ATOMS=2,4

METAD ...
  LABEL=metad
  ARG=d1,d2
  SIGMA=0.05
  HEIGHT=0.3
  PACE=200
  BIASFACTOR=10
... METAD

# The module of ITS must be declared behind the module of metadynamics
ITS_BIAS ...
  LABEL=its
# Set the label of metadynamics as the input bias
  BIAS=metad
  ARG=energy
  NREPLICA=100
  SIM_TEMP=300
  TEMP_MAX=370
  TEMP_MIN=270
  PACE=200
  RW_TEMP=280,300,320,340,360
... ITS_BIAS

PRINT ...
  ARG=d1,d2
  FILE=colvar.info.data
  STRIDE=100
... PRINT

PRINT ...
  ARG=energy,metad.*,its.*
  FILE=colvar.bias.data
  STRIDE=100
... PRINT
```

## S-IV  Reweighting of MetaITS

Here is an example to reweight the data from MetaITS:

```
d1: READ FILE=colvar.info.data VALUES=d1 IGNORE_TIME
d2: READ FILE=colvar.info.data VALUES=d2 IGNORE_TIME
metad: READ FILE=colvar.bias.data VALUES=metad.rbias IGNORE_TIME
its: READ FILE=colvar.bias.data VALUES=its.* IGNORE_TIME

rw280: REWEIGHT_BIAS ARG=metad.rbias,its.rbias_T280 TEMP=280
rw300: REWEIGHT_BIAS ARG= metad.rbias,its.rbias_T300 TEMP=300

HISTOGRAM ...
  LABEL=h280
  ARG=d1,d2
  GRID_MIN=0.0,0.0
  GRID_MAX=1.0,1.2
  GRID_BIN=100,120
  BANDWIDTH=0.2
  LOGWEIGHTS=rw280
... HISTOGRAM

HISTOGRAM ...
  LABEL=h300
  ARG=d1,d2
  GRID_MIN=0.0,0.0
  GRID_MAX=1.0,1.2
  GRID_BIN=100,120
  BANDWIDTH=0.2
  LOGWEIGHTS=rw300
... HISTOGRAM

DUMPGRID GRID=h280 FILE=histo.t280.data
DUMPGRID GRID=h300 FILE=histo.t300.data
```